\documentclass[useAMS,usenatbib]{mn2e}
\usepackage{graphicx}
\usepackage{epsfig}

\title[Companion stars of SNe Ia and single LMWDs]
{Companion stars of Type Ia supernovae and  single low-mass white dwarfs}
\author[B. Wang and Z. Han]
{B. Wang$^{\rm 1,2,3}$\thanks{E-mail:wangbo@ynao.ac.cn} and Z. Han$^{\rm 1,2}$  \\
$^1$National Astronomical Observatories/Yunnan Observatory, the
Chinese Academy of Sciences, Kunming 650011, China\\
$^2$Key Laboratory for the Structure and Evolution of Celestial
Objects, the Chinese Academy of Sciences, Kunming 650011, China\\
$^3$Graduate School of the Chinese Academy of Sciences, Beijing
100049, China}

\begin{document}
\pagerange{\pageref{firstpage}--\pageref{lastpage}} \pubyear{2010}
\maketitle

\label{firstpage}

\begin{abstract}
Recent investigations of the WD + MS channel of Type Ia supernovae
(SNe Ia) imply that this channel may be the main contribution to the
old population ($\ga$1\,Gyr) of SNe Ia. In the WD + MS channel, the
WD could accrete material from a main-sequence or a slightly evolved
star until it reaches the Chandrasekhar mass limit. The companions
in this channel would survive after SN explosion and show
distinguishing properties. In this Letter, based on SN Ia production
regions of the WD + MS channel and three formation channels of WD +
MS systems, we performed a detailed binary population synthesis
study to obtain the properties of the surviving companions. The
properties can be verified by future observations. We find that the
surviving companions of the old SNe Ia have a low mass, which
provides a possible way to explain the formation of the population
of single low-mass WDs ($<$0.45$\,M_{\odot}$).

\end{abstract}


\section{INTRODUCTION}\label{INTRODUCTION}
Type Ia supernovae (SNe Ia) appear to be good cosmological distance
indicators owing to their high luminosities and remarkable
uniformity, and have been applied successfully in determining
cosmological parameters (e.g. $\Omega$ and $\Lambda$; Riess et al.
1998; Perlmutter et al. 1999). However, several key issues related
to the nature of their progenitors and the physics of the explosion
mechanisms are still not well understood (Hillebrandt \& Niemeyer
2000; Wang et al. 2008; Podsiadlowski 2010), and no SN Ia progenitor
system before the explosion has been conclusively identified. These
uncertainties may raise doubts about the distance calibration which
is purely empirical and based on the SN Ia sample of the low
red-shift Universe ($z<0.05$; Phillips 1993).

It is widely accepted that SNe Ia are thermonuclear explosions of
carbon--oxygen white dwarfs (CO WDs) in binaries (for the review see
Nomoto, Iwamoto \& Kishimoto 1997). Over the past few decades, two
families of SN Ia progenitor models have been proposed, i.e. the
double-degenerate (DD) and single-degenerate (SD) models. In the DD
model, two CO WDs with a total mass larger than the Chandrasekhar
(Ch) mass limit may coalesce and then explode as an SN Ia (Iben \&
Tutukov 1984; Webbink 1984; Han 1998). Although it is suggested that
the DD model likely leads to an accretion-induced collapse rather
than to an SN Ia (Nomoto \& Iben 1985), it is still too early to
exclude the model as it may contribute to a few SNe Ia (Howell et
al. 2006; Pakmor et al. 2010). In the SD model, the companion is
probably a main-sequence (MS) or a slightly evolved star (WD + MS
channel), or a red-giant (RG) star (WD + RG channel) (e.g. Whelan \&
Iben 1973; Hachisu, Kato \& Nomoto 1996; Li $\&$ van den Heuvel
1997; Yungelson \& Livio 1998; Langer et al. 2000; Han $\&$
Podsiadlowski 2004, 2006; Chen $\&$ Li 2007; L\"{u} et al. 2009;
Wang, Li \& Han 2010, hereafter WLH10; Meng \& Yang 2010a; Wang \&
Han 2010a). Meanwhile, a CO WD may also accrete material from a He
star to increase its mass to the Ch mass limit (WD + He star
channel; Solheim \& Yungelson 2005; Wang et al. 2009a,b; Ruiter,
Belczynski \& Fryer 2009; Wang \& Han 2010b). An explosion following
the merger of two WDs would leave no remnant, while the companion in
the SD model would survive and potentially be identifiable. A
surviving companion in the SD model would evolve to a WD finally,
and Hansen (2003) suggested that the SD model could potentially
explain the properties of halo WDs, e.g. their space density and
ages. Note that there has been no conclusive proof yet that any
individual object is the surviving companion of an SN Ia. It would
be a promising method to test SN Ia progenitor models by identifying
their surviving companions.

By considering the effect of the thermal-viscous instability of
accretion disk on the evolution of WD binaries, WLH10 recently
enlarged the regions of the WD + MS channel for producing SNe Ia.
According to a detailed binary population synthesis (BPS) approach,
WLH10 found that this channel is effective for producing SNe Ia
($\sim$$1.8\times 10^{-3}\ {\rm yr}^{-1}$ in the Galaxy). This study
also implied that the WD + MS channel may be the main contribution
to the old population ($\ga$1\,Gyr) of SNe Ia. The companions in
this channel would survive and show distinguishing properties.

At present, the existence of a population of single low-mass
($<$0.45$\,M_{\odot}$) WDs (LMWDs) is supported by some observations
(e.g. Kilic et al. 2007a). (1) The first two LMWDs have been implied
by the work of Marsh et al. (1995). They carried out radial velocity
measurements of 7 LMWDs, and found that two of the WDs, WD1353+409
and WD1614+136 do not show any radial velocity variations. (2)
Maxted et al. (2000) also presented 14 LMWDs with no detectable
radial velocity variations from a larger radial velocity survey of
71 WDs. (3) A recent analysis of 348 H atmosphere WDs from the
Palomar Green (PG) survey has revealed 30 LMWDs (Liebert et al.
2005). Kilic et al. (2007a) concluded that 47\% of the PG LMWDs
searched for companions seem to be single. (4) The ESO SN Ia
Progenitor Survey (SPY) searched for radial velocity variations in
more than a thousand WDs using the Very Large Telescope (Napiwotzki
et al. 2001). Kilic et al. (2007a) also concluded that 15 of 26
LMWDs discovered in the SPY project do not show any radial velocity
variations, corresponding to a single LMWD fraction of 58\% (also
Napiwotzki et al. 2007).

The formation of the single LMWDs is still unclear. It is suggested
that the single LMWDs could be produced by single old metal-rich
stars which experience significant mass-loss prior to the He flash
(Kalirai et al. 2007; Kilic et al. 2007a). However, within the age
of the Universe, it is almost certainly impossible for the single
stars to produce WDs with mass close to $0.3\,M_{\odot}$, or even
some extremely LMWDs with mass as low as $0.2\,M_{\odot}$ (e.g.
Kilic et al. 2007a,b; Justham et al. 2009). Furthermore, the study
of initial-final mass relation for stars by Han et al. (1994)
implied that only LMWDs with masses $>$$0.4\,M_{\odot}$ might be
produced from such a single-star channel, even at high metallicity
(Meng, Chen \& Han 2008). Thus, it would be difficult to conclude
that single stars can produce LMWDs. Justham et al. (2009) recently
inferred an attractive formation channel for the single LMWDs, which
could have been formed in binaries where their companions have
exploded as SNe Ia. Note that Nelemans \& Tauris (1998) also
proposed an alternative scenario to form single LMWDs from a
solar-like star accompanied by a massive planet, or a brown dwarf,
in a relatively close orbit.

Han (2008) obtained many properties of the surviving companions of
the SNe Ia with intermediate delay times (100\,Myr$-$1\,Gyr; the
delay times of SNe Ia are defined as the time intervals between the
star formation and SN Ia explosion). Wang \& Han (2009) studied the
properties of the companions of the SNe Ia with short delay times
($\la$100\,Myr) from the WD + He star channel. The properties can be
verified by future observations (e.g. the masses, the spatial
velocities, the effective temperatures, the luminosities, the
surface gravities, etc; referring to Wang \& Han 2009). The purpose
of this Letter is to investigate the properties of the surviving
companions of the old SNe Ia from the WD + MS channel inferred in
WLH10, and to explore whether the surviving companions, later in
their evolution, could explain the existing population of single
LMWDs. In Section 2, we describe the BPS method for obtaining the
properties of the companions. The BPS results and discussion are
given in Section 3.

\section{Binary population synthesis}
In the WD + MS channel, the progenitor of an SN Ia is either a close
WD + MS or WD + subgiant system, which has most likely emerged from
the common envelope (CE) evolution of a binary involving a giant
star. The CE ejection is still an open problem. Similar to the work
of Wang et al. (2009b), we also use the standard energy equations
(Webbink 1984) to calculate the output of the CE phase. For this
prescription of the CE ejection, there are two highly uncertain
parameters, i.e. $\alpha_{\rm ce}$ and $\lambda$, where $\alpha_{\rm
ce}$ is the CE ejection efficiency, and $\lambda$ is a structure
parameter that depends on the evolutionary stage of the donor. As in
previous studies, we combine $\alpha_{\rm ce}$ and $\lambda$ into
one free parameter $\alpha_{\rm ce}\lambda$, and set it to be 0.5
(for details see Wang et al. 2009b).

To obtain the distributions of properties of the surviving
companions, we performed a Monte Carlo simulation in the BPS study.
In the simulation, by using the Hurley's rapid binary evolution code
(Hurley et al. 2000, 2002), we followed the evolution of
$1\times10^{\rm 7}$ sample binaries from the star formation to the
formation of the WD + MS systems according to three evolutionary
channels (i.e. the He star channel, the EAGB channel and the TPAGB
channel; the three channels here all refer to the WD progenitor's evolutionary phase when it encounters a Roche lobe
overflow (RLOF) event, for details see WLH10). If a binary evolves to a WD + MS
system, and if the system, at the onset of the RLOF phase, is located in the SN Ia production regions
(see Fig. 6a of WLH10) in the plane of ($\log P^{\rm i}$, $M_2^{\rm
i}$) for its $M_{\rm WD}^{\rm i}$, where $P^{\rm i}$, $M_2^{\rm i}$
and $M_{\rm WD}^{\rm i}$ are, respectively, the orbital period, the
secondary's mass, and the WD's mass of the WD + MS system at the
onset of the RLOF, we assume that an SN Ia is resulted, and the
properties of the WD + MS system at the moment of SN explosion are
obtained by interpolation in the three-dimensional grid ($M_{\rm
WD}^{\rm i}$, $M_2^{\rm i}$, $\log P^{\rm i}$) of the $\sim$2400
close WD + MS systems calculated in WLH10.

In the BPS study, the primordial binary samples are generated in the
Monte Carlo way. We adopted the following input for the simulation
(e.g. Wang et al. 2009b, 2010).

(1) The initial mass function (IMF) of Miller \& Scalo (1979) is
adopted.

(2) We take a constant mass-ratio ($q'$) distribution (Goldberg \&
Mazeh 1994),
\begin{equation}
n(q')=1, \hspace{2.cm} 0<q'\leq1,
\end{equation}
where $q'=M_{\rm 2}^{\rm p}/M_{\rm 1}^{\rm p}$.

(3) We assume that all stars are members of binaries and that the
distribution of separations is constant in $\log a$ for wide
binaries, where $a$ is separation and falls off smoothly at a small
separation
\begin{equation}
a\cdot n(a)=\left\{
 \begin{array}{lc}
 \alpha_{\rm sep}(a/a_{\rm 0})^{\rm m}, & a\leq a_{\rm 0},\\
\alpha_{\rm sep}, & a_{\rm 0}<a<a_{\rm 1},\\
\end{array}\right.
\end{equation}
where $\alpha_{\rm sep}\approx0.07$, $a_{\rm 0}=10\,R_{\odot}$,
$a_{\rm 1}=5.75\times 10^{\rm 6}\,R_{\odot}=0.13\,{\rm pc}$ and
$m\approx1.2$. This distribution implies that the numbers of wide
binaries per logarithmic interval are equal, and that about 50\% of
stellar systems have orbital periods less than 100\,yr (Han,
Podsiadlowski \& Eggleton 1995).

(4) A circular orbit is assumed for all binaries. The orbits of
semidetached binaries are generally circularized by tidal forces on
a timescale which is much smaller than the nuclear timescale.

(5) The star-formation rate (SFR) is taken to be constant over the
past 15\,Gyr.

\section{Results and discussion}

The simulation gives current-epoch distributions of many properties
of companions at the moment of SN explosion, e.g. the masses, the
orbital periods, the orbital separations, the orbital velocities,
the effective temperatures, the luminosities, the surface gravities,
the surface abundances, the mass-transfer rates, the mass-loss rates
of the optically thick stellar winds, etc. The simulation also shows
the initial parameters of the primordial binaries and the WD + MS
systems that lead to SNe Ia. Figs 1-5 are selected distributions
that may be helpful for identifying the surviving companions.

\begin{figure}
\includegraphics[width=5.4cm,angle=270]{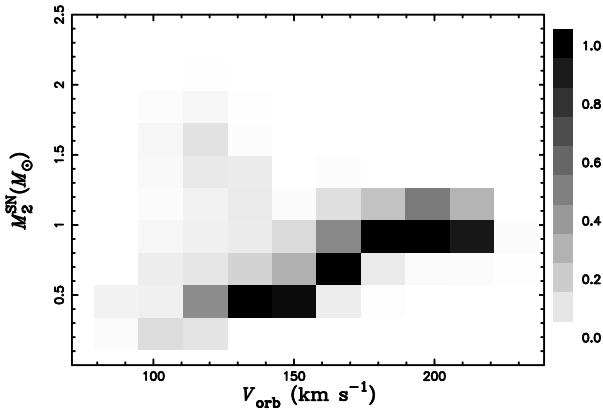}
\caption{Distribution of properties of the companions in the plane
of ($V_{\rm orb}$, $M_2^{\rm SN}$) at the current epoch, where
$V_{\rm orb}$ is the orbital velocity and $M_2^{\rm SN}$ the mass at
the moment of SN explosion.}
\end{figure}

Fig. 1 shows the distributions of the masses and the orbital
velocities of companions at the moment of SN explosion. In the
figure, the companion has an orbital velocity of
$\sim$80$-$220\,km/s for a corresponding mass of
$\sim$0.2$-$1.8$\,M_\odot$ at the moment of SN explosion. The
distributions correspond to the moment of SN explosion.  The
physical quantities depicted in the figure were calculated based on
the assumption that the companion (e.g. at the time when the WD
explodes) has not yet been affected by the explosion itself.  Since
SN explosion is expected to have a direct impact upon the surviving
companion, one would expect that the distributions would look
somewhat modified at a later time. For example, the SN ejecta will
interact with its companion. The companions will be stripped of some
mass (see the next paragraph) and receive a kick velocity that is
perpendicular to the orbital velocity. Marietta, Burrows \& Fryxell
(2000) presented several high-resolution two-dimensional numerical
simulations of the impact of SN Ia explosion with companions. The
study implied that this impact makes the companion in the WD + MS
channel receive a kick of 49$-$86\,km/s. With detailed stellar
models and realistic separations that were obtained from binary
evolution, Meng, Chen \& Han (2007) obtained a similar kick velocity
30$-$90\,km/s. Thus, a surviving companion has a space velocity
larger by $\sim$10\% than that in Fig. 1.

\begin{figure}
\includegraphics[width=8.5cm,angle=0]{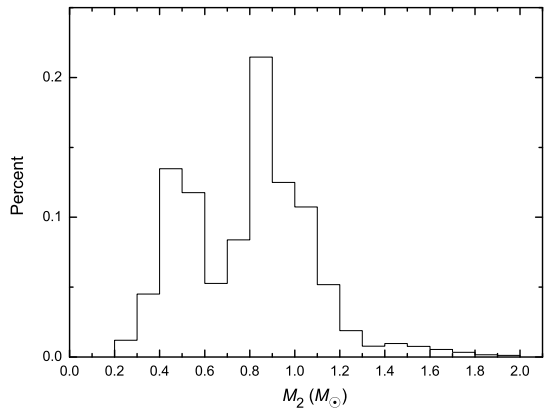}
\caption{Bimodal distribution of the mass for the companions at the
moment of SN explosion. }
\end{figure}

We find that the distribution of the mass for the companions is
bimodal (see Fig. 2). The left peak results from the companions of
SNe Ia with long delay times, while the right from the companions of
SNe Ia with intermediate delay times.  For SNe Ia with intermediate
delay times, SN Ia explosions occur when the companion is a MS or a
slightly evolved star. The study of Marietta, Burrows \& Fryxell
(2000) implied that the impact of the SN explosion makes the MS or
the slightly evolved companion lose a mass of
0.15$-$$0.17\,M_\odot$, while Meng, Chen \& Han (2007) obtained a
lower `stripped mass' $\sim$0.03$-$0.13$M_\odot$. A surviving
companion of SNe Ia with intermediate delay times therefore has a
mass lower by $\sim$0.1\,$M_\odot$. However, for the SNe Ia with
long delay times, SN Ia explosions occur when the companion evolves
to the RG stage. Marietta, Burrows \& Fryxell (2000) found that an
RG donor will lose almost its entire envelope (96\%$-$98\%) owing to
the impact of the SN explosion and leave only the core of the star.
Thus, the surviving companions of SNe Ia with long delay times will
contribute to the population of single LMWDs.

WLH10 inferred that the WD + RG channel of SNe Ia, in which the WD
begins to accrete material from an RG star, has a long delay time
from the star formation to SN explosion due to the low initial mass
($\la$1.5$\,M_{\odot}$) of the RG donors. So, all SNe Ia from this
channel also contribute to the old populations of SNe Ia, and the
companions have a low mass at the moment of SN explosion (see Fig. 2
of Wang \& Han 2010a), which also provides a possible pathway for
the formation of the single LMWDs. However, the Galactic SN Ia
birthrate from the WD + RG channel is low ($\sim$$3\times 10^{-5}\
{\rm yr}^{-1}$; WLH10) compared with observations, so in this study
we mainly focus on the investigation of surviving companions from
the WD + MS channel, which may be the main contribution to the
population of single LMWDs (note that the theoretical birthrate from
the WD + RG channel still contains many potential uncertainties).
Here, we emphasize that the surviving companions of the old SNe Ia
from the WD + MS and WD + RG channels provide a possible way for the
formation of the population of the single LMWDs. We also suggest
that the observed single LMWDs may provide evidence that at least
some SN Ia explosions have occurred with non-degenerate donors (such
as MS or RG donors).

The timescale from SN explosion to the formation of a WD from the
surviving companion is important in future observational tests,
which is related to the evolutionary phase of the surviving
companion. If the companion is in the MS stage at the moment of SN
explosion, the timescale mainly depends on the nuclear burning lifetime
of the star. We did a test for the evolution of a typical MS surviving
companion, which has the mass of 1.04$\,M_{\odot}$ and with central H
abundance 0.26 (its remaining MS lifetime is about 840\,Myr, and
its post MS lifetime is about
1300\,Myr). Thus, on average the nuclear burning lifetime of the MS surviving
companion is about 2\,Gyr. If the
companion is in the RG stage at the moment of SN explosion, the
timescale is mainly decided by the lifetime of the H-shell burning.
We also did a test for this case (e.g. a 0.42$\,M_{\odot}$ RG
companion star with 0.17$\,M_{\odot}$ He-core, the lifetime of the
H-shell burning is about 180\,Myr. Here, we ignore the thermal
equilibrium timescale from a He-core to a WD , since it is short
$\sim$$10^{6}$\,yr compared with the timescale of the H-shell
burning). Therefore, a low mass companion in the RG stage will
evolve to a WD more quickly than the massive companion in the MS
stage.

\begin{figure}
\includegraphics[width=5.4cm,angle=270]{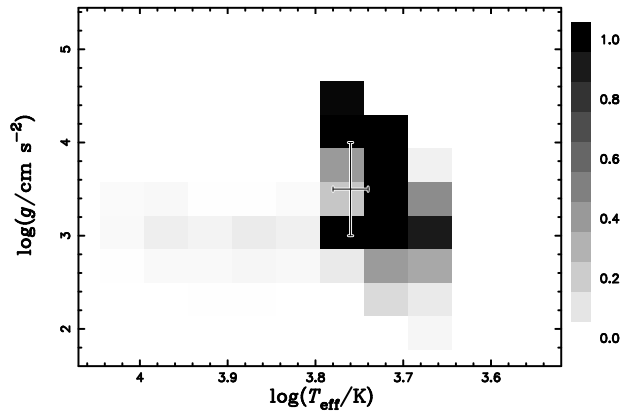}
\caption{Same as Fig. 1, but in the plane of ($\log T_{\rm eff}$,
$\log g$), where $T_{\rm eff}$ is the effective temperature of the
companions at the moment of SN explosion and $\log g$ the surface
gravity. The error bars denote the location of Tycho G
(Ruiz-Lapuente et al. 2004).}
\end{figure}

Fig. 3 represents the distributions of the effective temperatures
and the surface gravities of the companions at the moment of SN
explosion. Tycho G was taken as the surviving companion of Tycho's
SN by Ruiz-Lapuente et al. (2004). It has a space velocity of
$136\,{\rm km/s}$, more than 3 times the mean velocity of the stars
in the vicinity. Its surface gravity is $\log\, (g/{\rm cm}\, {\rm
s}^{-2})=3.5\pm 0.5$, while the effective temperature is $T_{\rm
eff}=5750\pm 250 {\rm K}$. The parameters are compatible with Figs 1
and 3. However, Fuhrmann (2005) argued that Tycho G might be a Milky
way thick-disk star that is coincidentally passing the vicinity of
the remnant of Tycho's SN. Ihara et al. (2007) recently also argued
that Tycho G may not be the companion of Tycho's SN, as the star did
not show any special properties in its spectrum. The surviving
companions of SNe Ia would expect to be contaminated by SN ejecta
and show some special characteristics (e.g. Marietta, Burrows \&
Fryxell 2000). Thus, whether Tycho G is the surviving companion of
Tycho's SN is still quite debatable.

\begin{figure}
\includegraphics[width=5.4cm,angle=270]{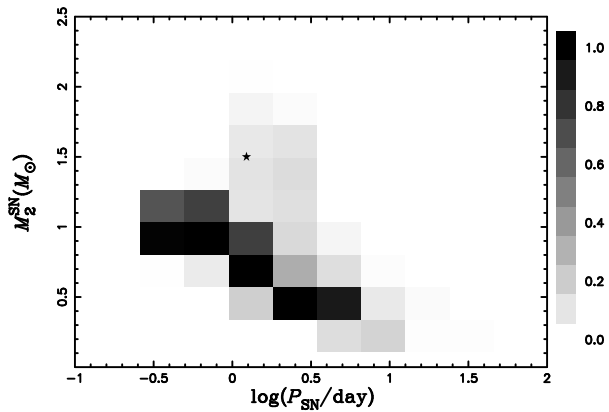}
\caption{Same as Fig. 1, but in the plane of ($\log P^{\rm SN}$,
$M_2^{\rm SN}$), where $P^{\rm SN}$is the orbital period at the
moment of SN explosion. The filled asterisk indicates the position
of a recurrent nova (U Sco).}
\end{figure}

Fig. 4 shows the distributions of orbital periods and secondary
masses of the WD + MS systems at the moment of SN explosion. The
orbital periods and secondary masses at this moment are basic input
parameters when one simulates the interaction between SN ejecta and
its companion. This figure may also help us to verify whether some
WD + MS systems observed could explode as SNe Ia. A recurrent nova
(U Sco) is indicated by a filled asterisk in the figure, in which
the WD mass is about $1.37\,M_{\odot}$ and its companion is a
1.5$\,M_{\odot}$ MS star (Hachisu et al. 2000). The orbital period
of the binary is 1.23\,d (Schaefer \& Ringwald 1995). Hachisu et al.
(2000) concluded that the WD could increase its mass until an SN Ia
explosion. If the WD explodes as an SN Ia eventually, the companion
would have a slightly smaller mass. The binary orbital period will
first decrease and then increase if the mass-ratio reverses. Then,
its final position in the figure will move to a lower mass than its
present one, and may enter into the most probable area for producing
SNe Ia. Thus, U Sco is likely to explode as an SN Ia (see also Meng
\& Yang 2010b).

If we assume that the companions co-rotate with their orbits, we can
obtain the distributions of their equatorial rotational velocities
(see Fig. 5). We see that the surviving companions are fast
rotators, so their spectral lines should be broadened noticeably.
The rotational velocity of companions from the WD + MS channel is in
the range of $\sim$10$-$140\,km/s, which is lower than that from the
WD + He star channel ($\sim$120$-$380\,km/s; Wang \& Han 2009). This
is because the WD + He star channel has shorter orbit periods than
that of the WD + MS channel at the moment of SN explosion.

\begin{figure}
\includegraphics[width=5.4cm,angle=270]{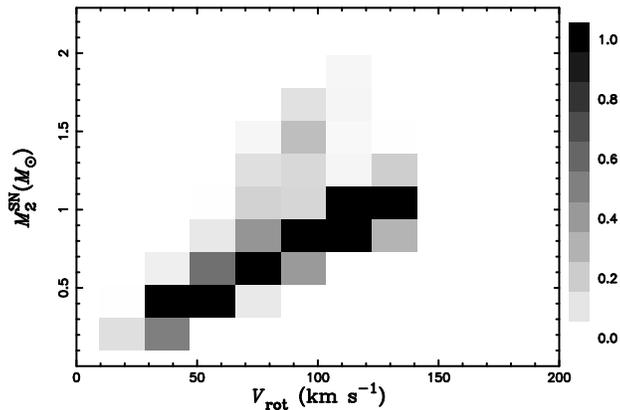}
\caption{Same as Fig. 1, but in the plane of ($V_{\rm rot}$,
$M_2^{\rm SN}$), where $V_{\rm rot}$ is the equatorial rotational
velocity of the companions at the moment of SN explosion.}
\end{figure}

The simulation in this Letter was made with $\alpha_{\rm ce}\lambda
=0.5$. If we adopt a higher value for $\alpha_{\rm ce}\lambda$ (e.g.
1.5), the birthrate of SNe Ia will be lower than the case of
$\alpha_{\rm ce}\lambda=0.5$ (the binaries emerged from the CE
ejections tend to have slightly closer orbits for $\alpha_{\rm
ce}\lambda=0.5$ and are more likely to be located in the SN Ia
production region). In addition, a high value of $\alpha_{\rm
ce}\lambda$ leads to a systematically later explosion time, i.e. the
delay time from the star formation to SN explosion will be longer.
This is because a high value of $\alpha_{\rm ce}\lambda$ leads to
wider WD + MS systems, and, as a consequence, it takes a longer time
for the companion to evolve to fill its Roche lobe. For a high value
of $\alpha_{\rm ce}\lambda$ (e.g. 1.5), the minimum delay time is
$\sim$360\,Myr, while the value is $\sim$280\,Myr for $\alpha_{\rm
ce}\lambda=0.5$.

\section*{Acknowledgments}
We thank the anonymous referee for valuable comments that helped us
to improve the paper. We also thank Stephen Justham for helpful
discussions and comments on the first version of this paper. This
work is supported by the National Natural Science Foundation of
China (Grant No. 10821061), the National Basic Research Program of
China (Grant No. 2007CB815406) and the Chinese Academy of Sciences
(Grant No. KJCX2-YW-T24).

\label{lastpage}
\end{document}